\documentclass[
 reprint,
 superscriptaddress,
 amsmath,amssymb,
 aps,
prl,
]{revtex4-2}

\usepackage{graphicx}
\usepackage{dcolumn}
\usepackage{physics}
\usepackage{hyperref}
\usepackage{amsmath}
\usepackage[T1]{fontenc}
\usepackage[utf8]{inputenc}
\usepackage{textgreek}
\hypersetup{
    colorlinks=true,
    linkcolor=blue,
    filecolor=magenta,
    urlcolor=blue,
    citecolor=blue,
}

\begin{document}

\title{Engineering Fractional Chern Insulators through Periodic Strain in Monolayer Graphene and Transition Metal Dichalcogenides}

\author{Yuchen Liu}
\affiliation{Kavli Institute for Theoretical Sciences, University of Chinese Academy of Sciences, Beijing 100190, China}

\author{Zheng Zhu}
\email{zhuzheng@ucas.ac.cn}
\affiliation{Kavli Institute for Theoretical Sciences, University of Chinese Academy of Sciences, Beijing 100190, China}

\date{\today}

\begin{abstract}
    We propose the realization of interaction-driven insulators in periodically strained monolayer graphene and transition metal dichalcogenides (TMDs). 
    Through extensive many-body exact diagonalization, we provide compelling evidence for various fractional Chern insulators (FCIs) in both strained monolayer graphene and TMDs, including the Laughlin states, Halperin states, and FCIs with tunable topological orders in Chern number $\left| \mathcal{C} \right|=2$ bands.
    We also discuss the relationship among band geometry, band filling and spin polarization.
    Notably, by examining both the entanglement spectrum and many-body Chern number, we reveal a state with Laughlin-like topological order emerging in the $\left| \mathcal{C} \right|=2$ band, which challenges the existing theoretical understanding of high Chern number (high-$\mathcal{C}$) FCIs.
    These findings suggest that periodically strained monolayer graphene and TMDs provide promising platforms for engineering fractional Chern insulators and underscore the need for further investigation into high-$ \mathcal{C} $ FCIs.  
\end{abstract}

\maketitle

\emph{Introduction.---} Fractional Chern insulators have been proposed to emerge in nearly flat Chern bands with fractional filling and no external magnetic field~\cite{Haldane2015, sheng_fractional_2011,neupert_fractional_2011,regnault_fractional_2011,tang_high-temperature_2011,qi_generic_2011,parameswaran_fractional_2013,bergholtz_topological_2013,wu_zoology_2012,lauchli_hierarchy_2013}.
However, the experimental realization of FCIs remains challenging due to stringent requirements on the band structure and geometry~\cite{roy_band_2014, jackson_geometric_2015}. Consequently, despite early theoretical predictions, FCIs have eluded experimental observation for a decade.

The recent discovery of correlated states in moiré materials~\cite{andrei2020graphene,andrei2021marvels,morales2024fractionalized} has significantly advanced the study of FCIs. By precisely adjusting the twist angle and gate voltage, the bandwidth and band geometry of these materials can be finely tuned, facilitating the realization of topologically nontrivial and strongly interacting bands. This has prompted numerous theoretical predictions of FCIs in moiré systems~\cite{wu_topological_2019,li_spontaneous_2021,devakul_magic_2021,yu_giant_2020,zhang_electronic_2021,crepel_anomalous_2023,reddy_fractional_2023,ledwith_fractional_2020,abouelkomsan_particle-hole_2020,wilhelm_interplay_2021,repellin_chern_2020}. In particular, the FCIs have been observed in both twisted bilayer of transition metal dichalcogenides (TMD)~\cite{zeng_thermodynamic_2023,kang_observation_2024,park_observation_2023,xu_observation_2023,cai_signatures_2023} and graphene-based moiré materials~\cite{lu_fractional_2024}.
These discoveries have sparked significant interest in understanding the emergent physics resulting from strong correlations and band topology in various moiré heterostructures.

By contrast, strain engineering offers another pathway to realizing topological flat bands with strong correlations. In two-dimensional materials, strain acts as a pseudomagnetic field (PMF)~\cite{fang_electronic_2018,vozmediano_gauge_2010,cortijo_electronic_2007,ochoa_emergent_2017,levy_strain-induced_2010}. Although creating a uniform PMF requires quadratic atomic displacement with distance~\cite{guinea_energy_2010,low_strain-induced_2010}, a more experimentally feasible scheme involving periodic strain engineering has been proposed to achieve topological flat bands for monolayer graphene~\cite{gao_untwisting_2023,phong_boundary_2022,milovanovic_band_2020,mahmud_topological_2023,wan_nearly_2023,zhai_supersymmetry_2024}. This setup could potentially be realized experimentally by thermal treatments~\cite{bao_controlled_2009,meng_hierarchy_2013}, mechanical manipulation~\cite{jiang_visualizing_2017,nemes-incze_preparing_2017,tomori_introducing_2011,lu_transforming_2012,reserbat-plantey_strain_2014,palinkas_moire_2016,pacakova_mastering_2017,zhang_electronic_2018,yang_strain_2021} or through the spontaneous buckling of graphene on substrates like NbSe$_2$~\cite{mao_evidence_2020}.

Motivated by the above, we conduct an extensive exact diagonalization (ED) study on periodically strained monolayer graphene (PSMG) and TMDs (PSTMDs), demonstrating that the periodic strain provides a feasible platform to engineer various FCIs, including the conventional single-component FCIs, the two-component Halperin FCIs and the high-$\mathcal{C}$ FCIs. Our findings illustrate that strain engineering offers a widely tunable way to investigate the relationship between band geometry, spin-polarization mechanisms, and the emergence of distinct topological orders.

For PSMG, we identify a wide range of strain parameters capable of producing topologically nontrivial nearly flat bands, characterized by Chern numbers $\left| \mathcal{C} \right|=1,2$.
Our focus lies in realizing spinful two-component FCIs in the $\left| \mathcal{C} \right|=1$ band, as well as spin-polarized high-$\mathcal{C}$ FCIs in $\left| \mathcal{C} \right|=2$ bands. 
Unlike $\left| \mathcal{C} \right|=1$ FCIs with conventional fractional quantum Hall (FQH) states as counterparts~\cite{qi_generic_2011,wu_gauge-fixed_2012,liu_fractional_2013}, high-$\mathcal{C}$ Chern band can not be mapped to a single Landau level.
Previous studies have sought to understand the topological order of high-$\mathcal{C}$ FCIs by decomposing the high-$\mathcal{C}$ bands into $\mathcal{C}$ coupled bands, each with Chern number $\left| \mathcal{C} \right|=1$~\cite{wu_bloch_2013,wu_haldane_2014,dong_many-body_2023}, which predicts a Halperin-like topological nature for high-$\mathcal{C}$ FCIs.
In our study, we provide evidence from particle-cut entanglement spectrum (PES) and many-body Chern number calculations that a Halperin-like state is indeed present in the $\left| \mathcal{C} \right|=2$ band. However,  under different strain parameters, we also observe a Laughlin-like state in another $\left| \mathcal{C} \right|=2$ band with Hall conductivity $|\sigma_{xy}| \neq \nu \mathcal{C}_{\text{band}} e^2/h$, challenging existing theoretical understandings of high-$\mathcal{C}$ FCIs.
We offer an alternative explanation for the emergence of this Laughlin-like state in the $\left| \mathcal{C} \right|=2$ band, underscoring the novel insights of our findings.

We also investigate the potential realization of FCIs in periodically strained TMDs (PSTMDs). By modeling this system as massive fermions in a periodic PMF, we find that the valence band of PSTMDs can be tuned to form a nearly ideal topological flat band with realistic strain strengths. Utilizing ED, we find that $\nu = 1/3$ Laughlin FCI can be stabilized in PSTMDs, potentially simplifying experimental realization as it requires no finely tuned background scalar potential.

\begin{figure}[tbp]
    \centering
    \includegraphics[width=0.5\textwidth]{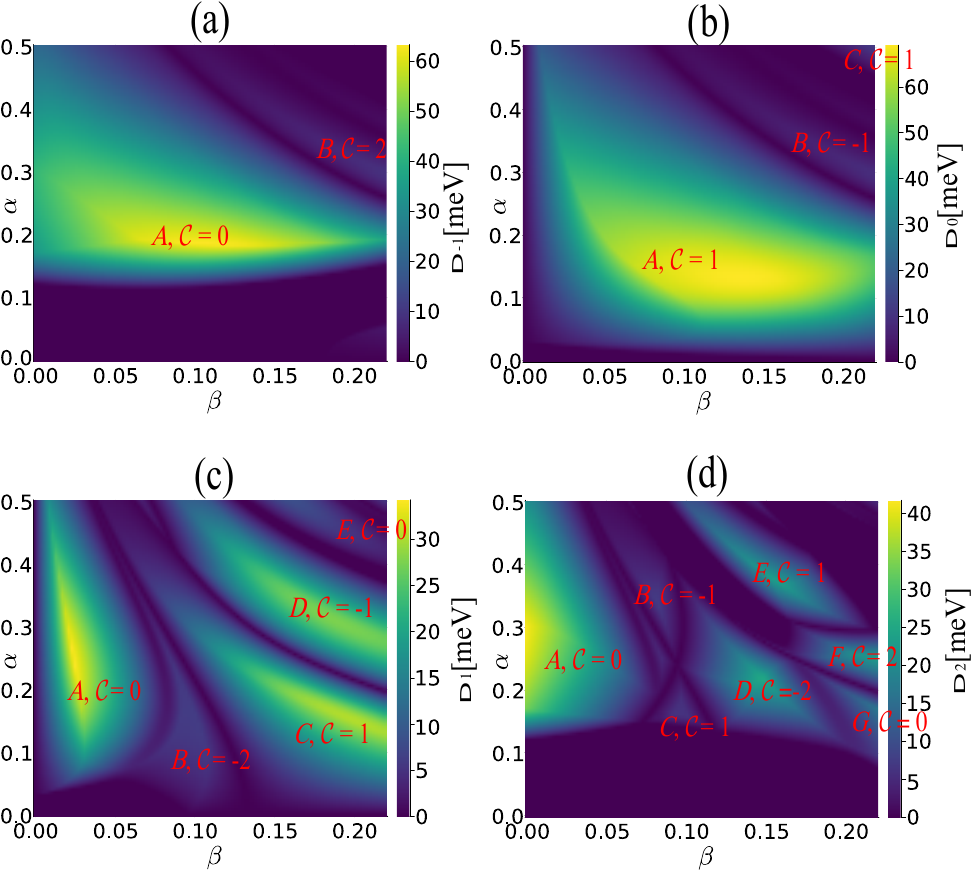}
    \caption{The indirect gap $\Delta_i$ of each band of PSMG as a function of dimensionless strain parameters $\alpha$ and $\beta$. Panels (a-d) correspond to band indices $i=-1,0,1,2$, respectively. The Chern number $\mathcal{C}$ of each band is indicated in the main regions where the bands are isolated.  These regions are labeled as $X_{\left(i\right)}$, where $i$ is the band index and letter $X$ represents different regions.}
    \label{fig:PSMG_band_gap}
\end{figure}

\emph{Model setup.---} We begin with modeling the PSMG and PSTMDs. The periodic strain acts as the PMF given by
$\mathbf{B}=B_0\mathbf{e}_z \sum_{l=0}^{5} e^{i \mathbf{G}_l\cdot\mathbf{r}}$, where $\mathbf{G}_{l}$ are constructed from $\mathbf{G}_0=(4\pi /\sqrt{3}{L_M},0)$ by rotations $\mathbf{G}_{l}=R_{\pi l/3}\mathbf{G}_{0}$, and $L_\mathrm{M}$ denotes the period of PMF. For PSMG, to obtain isolated flat bands, we additionally consider a periodical scalar potential 
$V(\mathbf{r})=V_0\sum_{l=0}^{5}e^{{i\mathbf{G}_{l}}\cdot\mathbf{r}}$ with the same periodicity to gap out the Dirac point~\cite{gao_untwisting_2023,Low2011}. By contrast, such scalar potential is not necessary in PSTMDs.

The continuum model of PSMG is
\begin{equation}
    \mathcal{H}_{\mathrm{PSMG}}={E_0}\left[\left(\vb*{k}+\alpha \vb*{\mathcal{A}} \right)\cdot\vb*{\sigma }+\beta V\left( {\vb*{r}}\right)\right],
\end{equation}
where $\vb*{\mathcal{A}}$ and $V$ are dimensionless gauge and scalar potentials, and $\alpha$ and $\beta$ are dimensionless strain parameters describing the strength of the PMF and scalar potential, respectively. $E_0=\hbar v_\mathrm{F} |\mathbf{G_0}|\approx 270\text{meV}$ is an energy scale determined by $L_\mathrm{M}$ and the Fermi velocity. 
We calculate the indirect gap $\Delta_i$ of four bands (including second and first valence bands, first and second conduction bands, $i$ refers to band index $-1,0,1,2$ respectively) near the charge neutrality point (CNP) as a function of $\alpha$ and $\beta$ in an experimentally accessible region. 
We observe various parameter regions of each band with significant $\Delta_i$ [see Figs.~\ref{fig:PSMG_band_gap}(a-d)],  
in which the isolated bands have Chern numbers $\left| \mathcal{C} \right|=0,1,2$. Notably, the bandwidth and fluctuations of both Berry curvature and Fubini-Study metric are minimal in specific regions (see supplementary~\cite{SM}),   
providing opportunities for realizing various FCIs in PSMG.

\begin{figure}[tbp]
    \centering
    \includegraphics[width=0.5\textwidth]{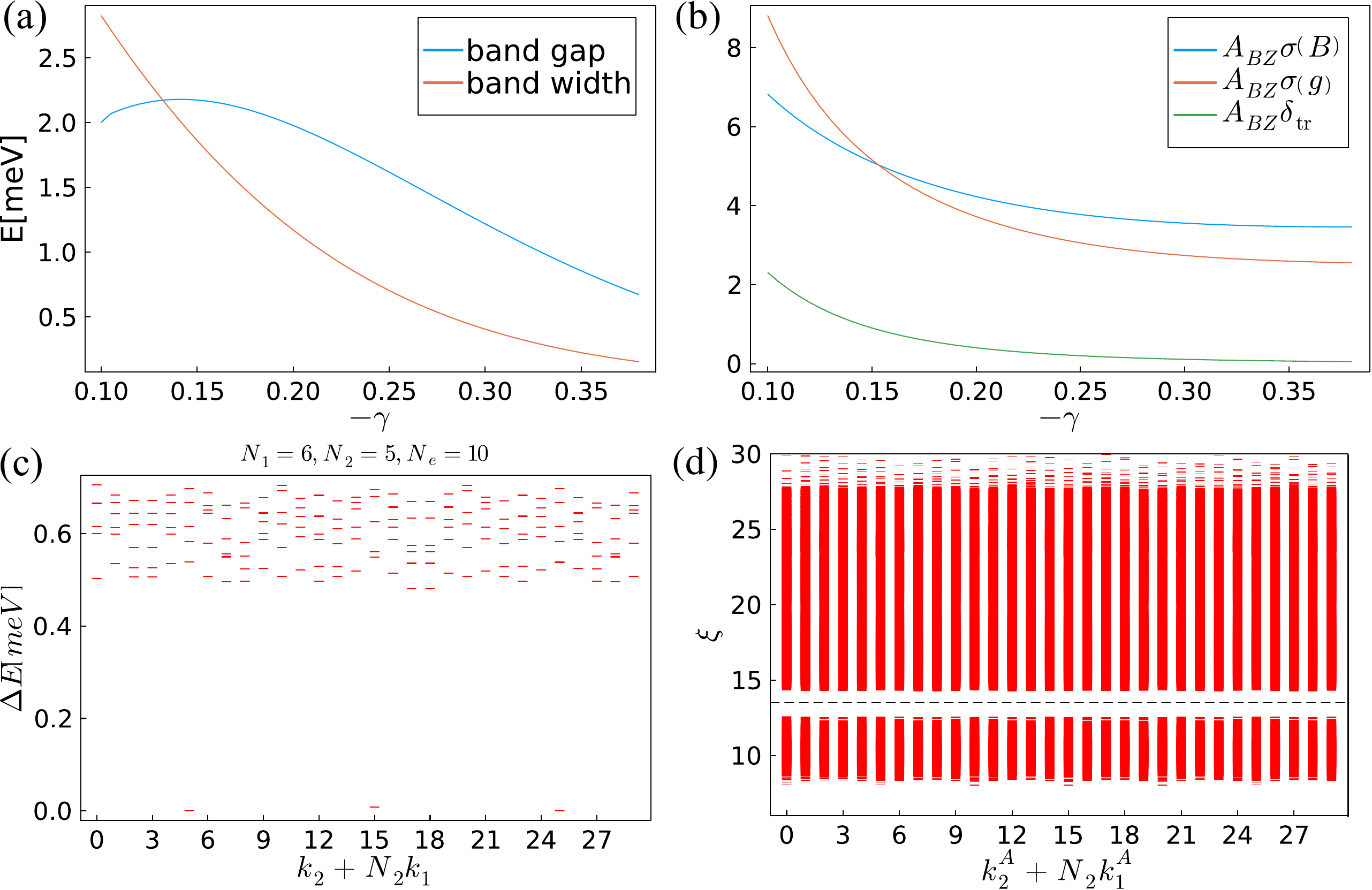}
    \caption{Evidence of $\nu=1/3$ Laughlin FCIs in PSTMD. 
    (a) Bandwidth and indirect band gap of the valence band as a function of $\gamma$. 
    (b) Fluctuation of Berry curvature and Fubini-Study metric of valence band of PSTMD versus $\gamma$.
At $\gamma = -0.25$ and $\nu = 1/3$ on a $6 \times 5$ system, panels (c-d) show the low-lying energy spectra (c) and particle entanglement spectrum (d)  for $N_A = 5$.}

    \label{fig:PSTMD}
\end{figure}

Unlike the monolayer graphene, the spin and valley are locked in TMDs. The PSTMDs can be modeled as massive Dirac fermions in a periodic PMF, with the continuum model given by  
\begin{equation}
    {{\mathcal{H}}_{\mathrm{PSTMD}}}=-{{E}_{0}}'{{\Pi }_{+}}{{\Pi }_{-}},
\end{equation}
where ${{\Pi }_{+}}=\Pi _{-}^{*}=\mathcal{K}-\gamma \mathcal{A}$, $\mathcal{K}=\left( {{k}_{x}}+i{{k}_{y}} \right)/\left| {{\mathbf{G}}_{0}} \right|$ is dimensionless wave vector, $\gamma$ is the dimensionless PMF strength, and ${{E}_{0}}'={{\hbar }^{2}}{{\left| {{\mathbf{G}}_{0}} \right|}^{2}}/2{{m}^{*}}$ is an energy scale determined by both $L_\mathrm{M}$ and the effective mass $m^*$ of the valence-band electrons. 
This model produces topologically nontrivial minibands, taking $\text{MoS}_2$ as an example, with maximum strain strength less than 5\% and ${{L}_{M}}\sim 15\text{nm}$, $\left| \gamma  \right|$ can reach at most 0.38 according to the parameters in Ref.~\cite{fang_electronic_2018}. For $0<-\gamma <0.38$, we find the valence band exhibits $\left| \mathcal{C} \right|=1$. In this parameter region, we also find that the bandwidth and fluctuations of both Berry curvature and  Fubini-Study metric of valence band are significantly reduced for sufficiently large $\left| \gamma  \right|$ [see Fig.~\ref{fig:PSTMD} below], offering opportunities for hosting FCIs in PSTMDs.

We consider electrons interacting via the screened Coulomb potential in both cases and choose the dual gate potential $\tilde{V}\left( {\vb*{q}} \right)=\left( {{e}^{2}}/2S{{\epsilon }_{r}}{{\epsilon }_{0}}q \right)\tanh \left( q\lambda /2 \right)$, where $S$ denotes the system area, $\epsilon_0$ ($\epsilon_r$) is the vacuum permittivity (relative permittivity),   
and $\lambda$ is the screening length. We set $\epsilon_r=4$ and $\epsilon_r=15$ for PSMG and PSTMDs, respectively, which are typical values for graphene and TMDs~\cite{abouelkomsan_particle-hole_2020,dong_composite_2023,reddy_fractional_2023}.
When a band $p$ is isolated from other bands, 
the projected total Hamiltonian to this active band is given by
\begin{equation}
    \begin{split}
    {{H}_{\text{proj}}}=\sum\limits_{\mathbf{k}}{{{\varepsilon }_{\mathbf{k}}}c_{\mathbf{k}}^{\dagger }{{c}_{\mathbf{k}}}}+\sum\limits_{\left\{ {{\mathbf{k}}_{i}} \right\}}{{{V}_{{{\mathbf{k}}_{1}}{{\mathbf{k}}_{2}}{{\mathbf{k}}_{3}}{{\mathbf{k}}_{4}}}}c_{{{\mathbf{k}}_{1}}}^{\dagger }c_{{{\mathbf{k}}_{2}}}^{\dagger }{{c}_{{{\mathbf{k}}_{3}}}}{{c}_{{{\mathbf{k}}_{4}}}}}
    \end{split}
    \label{eq:projected_continuum}
\end{equation}
where $\varepsilon_{\mathbf{k}}$ is the dispersion of the active band, $c_{\mathbf{k}}^{\dagger }$ creates an electron with momentum $\mathbf{k}$ in the active band, and $V_{{\mathbf{k}}_{1}{\mathbf{k}}_{2}{\mathbf{k}}_{3}{\mathbf{k}}_{4}}$ can be derived by band projection~\cite{abouelkomsan_particle-hole_2020,liu_recent_2023}.

\begin{figure*}[tbp]
    \centering 
    \includegraphics[width=1.0\textwidth]{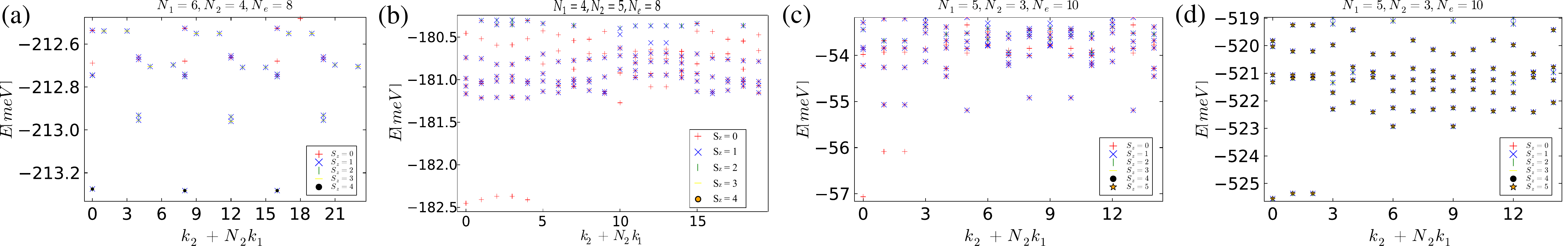}
    \caption{Energy spectra for different $S_z$ sectors of PSMG at $\nu=1/3,2/5,2/3$. 
    (a-c) $\nu=1/3,2/5,2/3$ for $\alpha=0.4,\beta=0.068$ in region $A_{(0)}$.
    (d) $\nu=2/3$ for $\alpha=0.29,\beta=0.22$ in region $B_{(0)}$. }
    \label{fig:PSMG_C_1}
\end{figure*}
 
\emph{FCIs in PSMG $\left| \mathcal{C} \right|=1$ bands.---}  
Based on extensive numerical and experimental results for spinful and valleyful fractional Chern bands~\cite{repellin_chern_2020, serlin_intrinsic_2020, burg_correlated_2019, sheng_quantum_2024}, we assume valley polarization for both PSMG and PSTMDs, while allowing free spin degrees of freedom in PSMG.

Previous studies have shown that the $\nu=1/3$ Laughlin FCI can be realized in PSMG with both valley and spin polarization~\cite{gao_untwisting_2023}. Relaxing the spin polarization, we calculate the energy spectra of PSMG at various fillings, including $\nu=1/3$, $2/5$, and $2/3$, at strain parameters $\alpha=0.4$ and $\beta=0.068$ in region $A_{(0)}$, which hosts a nearly ideal $\mathcal{C} =1$ band. We find that the ground state at $\nu=1/3$ is spin-polarized with a three-fold quasi-degeneracy [see Fig.~\ref{fig:PSMG_C_1}(a)], 
characterizing the $\nu=1/3$ Laughlin state~\cite{regnault_fractional_2011,bernevig_emergent_2012}. To confirm its topological nature, we further examine the PES and spectral flow under magnetic flux insertion in the supplementary material~\cite{SM}.

As the filling increases to $\nu=2/5$ and $\nu=2/3$ in the same Chern band, the ground states become spin-unpolarized with five-fold and three-fold quasi-degeneracies, respectively, suggesting the realization of the Halperin (332) and (112) states [see Fig.~\ref{fig:PSMG_C_1}(b-c)]. We also confirm their robustness against system sizes~\cite{SM}. Given that FCIs in a nearly ideal $\mathcal{C}=1$ band mimic weak-Zeeman fractional quantum Hall systems, the transition from a spin-polarized to a spin-unpolarized ground state with increasing filling is understandable~\cite{SM}.
However, by tuning the band-geometry fluctuations through straining fields, we observe 
that increasing fluctuations can induce spin polarization at fixed fillings that would otherwise remain spin-unpolarized. 
For instance, in region $B_{(0)}$ with strain parameters $\alpha=0.29$ and $\beta=0.22$, where band geometry fluctuations are more pronounced than $A_{(0)}$~\cite{SM}, we find a spin-polarized ground state at $\nu=2/3$ [see Fig.~\ref{fig:PSMG_C_1}(d)].
Notably, the tendency for spin polarization by larger band-geometry fluctuations aligns with previous studies on twisted bilayer graphene~\cite{repellin_chern_2020}, suggesting a potentially universal mechanism driving spin polarization at specific fillings.

\begin{figure}[tbp]
    \centering
    \includegraphics[width=0.5\textwidth]{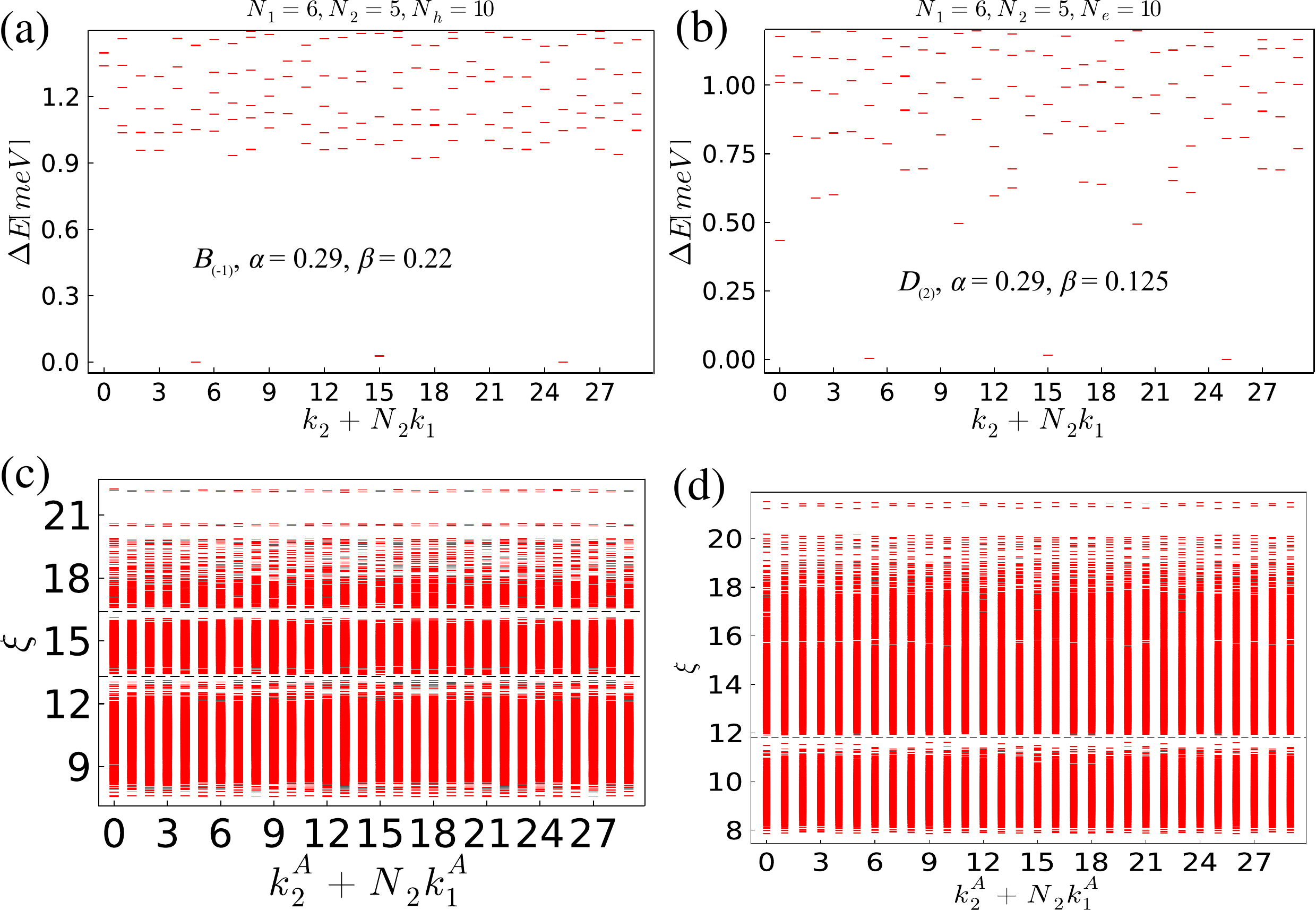}

    \caption{Evidence of $\mathcal{C}=2, \nu_{h/e}=1/3$ high-$\mathcal{C}$ FCIs in PSMG for regions $B_{(-1)}$ and $D_{(2)}$.
    (a-b) The low-lying energy spectra on a $6\times5$ system with $N_h=10$ holes (a) and $N_e=10$ electrons (b).  
    (c) and (d) show the particle entanglement spectrum (PES) for ground states in (a) and (b) with $N_A=4$, respectively.}
    \label{fig:PSMG_high_C}
\end{figure}

\emph{High Chern number FCIs.---} We also observe high-$\mathcal{C}$ regions in PSMG [see Fig.~\ref{fig:PSMG_band_gap}], which may host high-$\mathcal{C}$ FCIs with exotic topological order~\cite{wang_fractional_2012,liu_fractional_2012,yang_topological_2012,wang_tunable_2013}.
It has been demonstrated that ideal $\left|\mathcal{C}\right|=n$ bands can be constructed by twisting two sheets of Bernal-stacked $n$ graphene layers in the chiral limit~\cite{wang_hierarchy_2022,ledwith_family_2022}. 
For $n=2$, previous theoretical study interprets the $\nu=1/3$ high-$\mathcal{C}$ FCI in this system as a Halperin-like state exhibiting emergent $SU\left(2\right)$ symmetry, supported by PES counting~\cite{dong_many-body_2023,liu_gate-tunable_2021}.

In contrast, we present compelling evidence for realizing high-$\mathcal{C}$ FCIs with distinct topological orders in PSMG. Through extensive ED calculations, we identify that regions $B_{\left(-1\right)}$ and $D_{\left(2\right)}$ in PSMG can host $\left|\mathcal{C}\right|=2$ FCIs with finely tuned strain parameters, and these $\left|\mathcal{C}\right|=2$ FCIs exhibit distinct topological orders: one is Halperin-like, while the other is Laughlin-like.

In region $B_{\left(-1\right)}$, we find spin-polarized ground states at $\nu=2/3$~\cite{SM}. To facilitate comparison with the $\nu=1/3$ Laughlin state, we perform ED calculations in hole picture $\nu_h=1/3$. As shown in Fig. \ref{fig:PSMG_high_C}(a), there is a clear three-fold ground-state degeneracy in the energy spectrum for $\alpha=0.29$ and $\beta=0.22$. In the PES, we observe two clear entanglement gaps with 17250 levels below the lower gap and 24840 levels below the higher gap for $N_A=4$ [see Fig.~\ref{fig:PSMG_high_C}(c)], consistent with the PES counting of Halperin (112) state. 

In region $D_{\left(2\right)}$, we also identify a three-fold degeneracy in the low-lying energy spectrum at $\nu=1/3$ for $\alpha=0.29$ and $\beta=0.125$, and confirm the ground state is spin-polarized~\cite{SM}. However, the $N_A=4$ PES of the ground state shows only one entanglement gap with 9975 levels below it, aligning with the PES counting of the $\nu=1/3$ Laughlin state.
Since the low-lying entanglement spectrum serves as a ``fingerprint" for identifying topological order~\cite{li_entanglement_2008,sterdyniak_extracting_2011}, the distinct PES counting of these two regions strongly suggests that the $\left|\mathcal{C}\right|=2$ FCIs in these two regions exhibit different topological orders, with one being Halperin-like and the other Laughlin-like. 

\begin{figure}[tbp]
    \centering
    \includegraphics[width=0.5\textwidth]{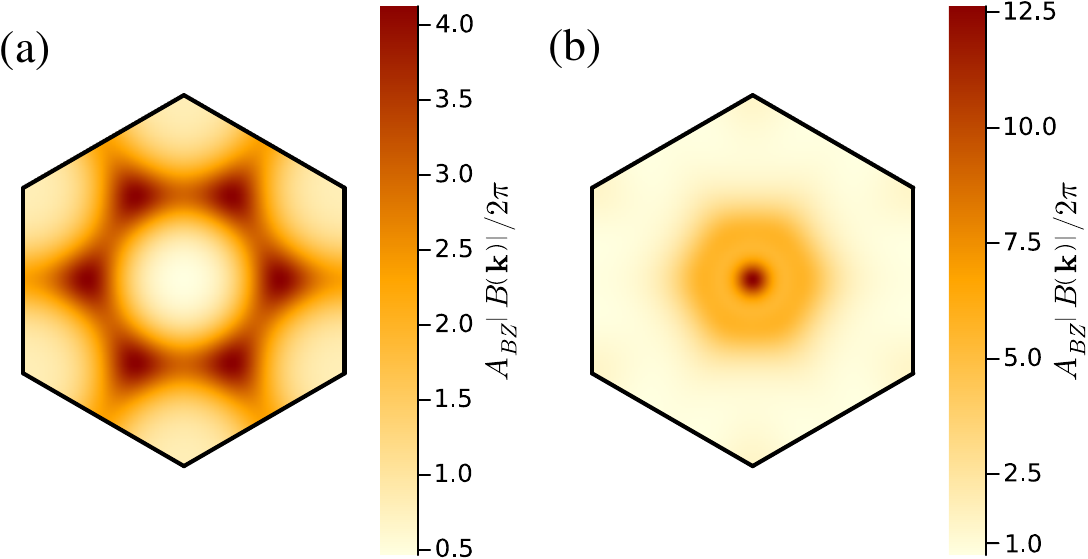}

    \caption{Overview of the single-particle band Berry curvature distribution for two different parameter points in PSMG. Panels (a-b) show the Berry curvature distribution for $\alpha=0.29,\beta=0.22$ in region $B_{(-1)}$ and $\alpha=0.29,\beta=0.125$ in region $D_{(2)}$, respectively. $A_{BZ}$ is the area of the Brillouin zone.}
    \label{fig:berry_distribution}
\end{figure}

To further confirm the topological nature of these high-$\mathcal{C}$ FCIs, we calculate the many-body Chern numbers of these two cases, which are integral invariant of many-body wave function over twist boundary condition:
\begin{equation}
    {{\mathcal{C}}_{j}}=\frac{i}{2\pi }\int{\text{d}{{\Phi }_{1}}\text{d}{{\Phi }_{2}}\left\{ \left\langle  {{\partial }_{{{\Phi }_{1}}}}{{\psi }_{j}} | {{\partial }_{{{\Phi }_{2}}}}{{\psi }_{j}} \right\rangle -c.c. \right\},}
    \label{eq:many-body_chern}
\end{equation}
where $j=1,2,3$ for three quasi-degenerate ground states, $\left|{{\psi }_{j}}\right\rangle$ is the many-body ground state, and $\Phi_1$ and $\Phi_2$ are the twist boundary phases in two directions. In a $10\times 10$ mesh, we find that the many-body Chern number of the ground states in Fig. \ref{fig:PSMG_high_C}(a) is $\sum_{j}{{{\mathcal{C}}_{j}}}=-2$ consistent with the Halperin (112) state, while the many-body Chern number of the ground states in Fig. \ref{fig:PSMG_high_C}(b) is $\sum_{j}{{{\mathcal{C}}_{j}}}=-1$, consistent with the $1/3$ Laughlin state.
This leads to an intriguing experimental consequence: quantized Hall conductance of above Laughlin-like state is unexpected from the filling factor and band Chern number, i.e., $\left| {{\sigma }_{xy}} \right|=1/3\ne\nu {{\mathcal{C}}_{\text{band}}}$ in unit of $e^2/h$.
We remark that this inequality in $\mathcal{C}=2$ band of PSMG is reminiscent of the recently discovered quantum anomalous Hall crystal in $\mathcal{C}=1$ band of twisted TMDs, which exhibits integer $\left| {{\sigma }_{xy}} \right|=1$ at $\nu=1/2$~\cite{sheng_quantum_2024}. But the Laughlin-like state we propose here does not break the translation symmetry of the original lattice in contrast~\cite{SM}.

To understand the unexpected Laughlin-like states in $\left|\mathcal{C}\right|=2$ bands, we propose an alternative mechanism based on the single-particle band Berry curvature distributions $B({\mathbf{k}})$ at strain parameters that host these two distinct states. As shown in Figs.~\ref{fig:berry_distribution}, 
unlike the smooth $B({\mathbf{k}})$ for the Halperin-like state [see Fig.~\ref{fig:berry_distribution}(a)], the Laughlin-like state [see Fig.~\ref{fig:berry_distribution}(b)] exhibits a significant characteristic: a nearly uniform  
$B({\mathbf{k}})$ across the Brillouin zone (with a value close to $2\pi/A_{BZ}$), with pronounced peaks localized around momentum $\Gamma$.
To gain a qualitative understanding, we idealize $B({\mathbf{k}}) =2\pi \left( 1/{{A}_{BZ}}+\delta \left( \mathbf{k} \right) \right)$. Under a long-wavelength approximation and by expanding to second order, we derive the relationship of the projected density operators~\cite{roy_band_2014}
\begin{equation}
    \begin{aligned}
        \left[ {{{\bar{\rho }}}_{{{\mathbf{q}}_{1}}}},{{{\bar{\rho }}}_{{{\mathbf{q}}_{2}}}} \right]&=i{{\mathbf{q}}_{1}}\wedge {{\mathbf{q}}_{2}} \\ 
        & \times \sum\limits_{\mathbf{k}}{{{B}_{\mathbf{k}}}u_{\mathbf{k}-{{\mathbf{q}}_{1}}-{{\mathbf{q}}_{2}}}^{\dagger }{{u}_{\mathbf{k}}}\left| \mathbf{k}-{{\mathbf{q}}_{1}}-{{\mathbf{q}}_{2}} \right\rangle \langle \mathbf{k}|}+\mathcal{O}\left( {{q}^{3}} \right),  \\ 
    \end{aligned}
    \label{eq:commutator}
\end{equation}
which corresponds to an approximate Girvin, MacDonald, and Platzman (GMP) algebra~\cite{GMP1,GMP2} modified by a single deviation term. This deviation arises solely from scatterings linked to the single-particle state at $\mathbf{k}=0$, and becomes negligible in the thermodynamic limit. This approximate GMP algebra in $\left|\mathcal{C}\right|=2$ bands explains the emergence of Laughlin-like states, which may be further stabilized when the $\Gamma$ point hosts higher band energy, thereby suppressing the corresponding scatterings. We also examine the ground-state momentum occupation $n\left( \mathbf{q} \right)=\left\langle c_{\mathbf{q}}^{\dagger }{{c}_{\mathbf{q}}} \right\rangle $ and find vanishing value at $\Gamma$, further supporting our explanations~\cite{SM}. 
This mechanism likely applies to other high-$\mathcal{C}$ FCIs characterized by a nearly uniform background Berry curvature across the Brillouin zone but with localized peaks in specific regions.

\emph{FCIs in PSTMDs.---} For PSTMDs, we calculate the bandwidth and indirect gap of the valence band as functions of the dimensionless PMF strength $\gamma$. As shown in Fig.~\ref{fig:PSTMD} (a-b). We select $\gamma=-0.25$ to ensure a small deviation from the ideal flat band condition while maintaining a significant band gap. Thus we focus on $\gamma=-0.25$ and choose $\lambda=15\text{nm}$ to study the Hamiltonian \eqref{eq:projected_continuum} by ED. At $\nu=1/3$,  we identify three-fold ground-state degeneracies in low-lying energy spectra for relatively large systems, as shown in Fig. \ref{fig:PSTMD}(c). The PES of the ground-state manifold shows 23256 levels below the entanglement gap for $N_A = 5$[see Fig.~\ref{fig:PSTMD}(d)], further demonstrating its topological order. These results provide compelling evidence for realizing the $\nu=1/3$ Laughlin state in PSTMDs. Notably, unlike the PSMG, no additional scalar potential is applied in PSTMDs, rendering it easier to achieve FCIs experimentally.

\emph{Summary.---} 
In this work, we analyze the band structure and geometry of periodically strained monolayer graphene and TMDs and conduct an extensive ED study of the single-band-projected models of these systems. Our results suggest the feasibility of engineering various FCIs through periodic strain, including the single-component FCIs in  $\mathcal{|C|}=1$ Chern bands, the two-component Halperin (112),(332) FCIs, and the high-$\mathcal{C}$ FCIs in $\mathcal{|C|}>1$ Chern bands. 
Our findings demonstrate that strain engineering provides an ideal platform to examine the relationship between the band geometry, spin-polarization mechanism and topological order origins.
Notably, we reveal that the $\mathcal{|C|}=2$ band in PSMG can host a Laughlin-like state. This finding challenges previous theoretical understandings of high-$\mathcal{C}$ FCIs and yields intriguing experimental implications, such as a quantized Hall conductance that deviates from the expected relation ${{\sigma }_{xy}}\ne \nu {{\mathcal{C}}_{\text{band}}}{{e}^{2}}/h$. Furthermore, we argue that this state is stabilized by appropriate Berry curvature distribution and band dispersion, contrasting with conventional requirements that necessitate small Berry curvature fluctuations and flat bands for FCI stabilization.
From an experimental perspective, implementing the necessary periodic scalar potential is required in PSMG, whereas it is not needed in PSTMDs. 
We anticipate that future experimental efforts will validate our theoretical predictions and further explore the rich physics of FCIs in strained systems. 
 
\begin{acknowledgments}
We acknowledge useful discussions with Hanwen Yang and Zhao Liu.
This work is supported by the National Natural Science Foundation of China (Grant No.12074375) and the Fundamental Research Funds for the Central Universities.
\end{acknowledgments}

\bibliography{ref}

\clearpage
\newpage

\begin{widetext}
\begin{center}
	\textbf{\large Supplemental Material}
\end{center}
\setcounter{equation}{0}
\setcounter{figure}{0}
\setcounter{table}{0}
\setcounter{section}{0}
\setcounter{page}{1}
\makeatletter
\renewcommand{\theequation}{S\arabic{equation}}
\renewcommand{\thefigure}{S\arabic{figure}}
\renewcommand{\bibnumfmt}[1]{[S#1]}
\renewcommand{\citenumfont}[1]{S#1}

\section{BAND STRUCTURE AND BAND GEOMETRY OF PSMG}
\label{sec:band_structure}
This section provides an overview of band structure and band geometry at different regions of PSMG that are not included
in the main text. We show the band width of each band of PSMG as a function of dimensionless parameters $\alpha$ and $\beta$ in [Fig. \ref{fig:PSMG_band_width}]. 
Remarkably, we find that the band width of these bands can be significantly small in some regions of the parameter space, offering a promising platform for the realization of interaction-driven states.
\begin{figure}[htbp]
    \centering
    \includegraphics[width=1.0\textwidth]{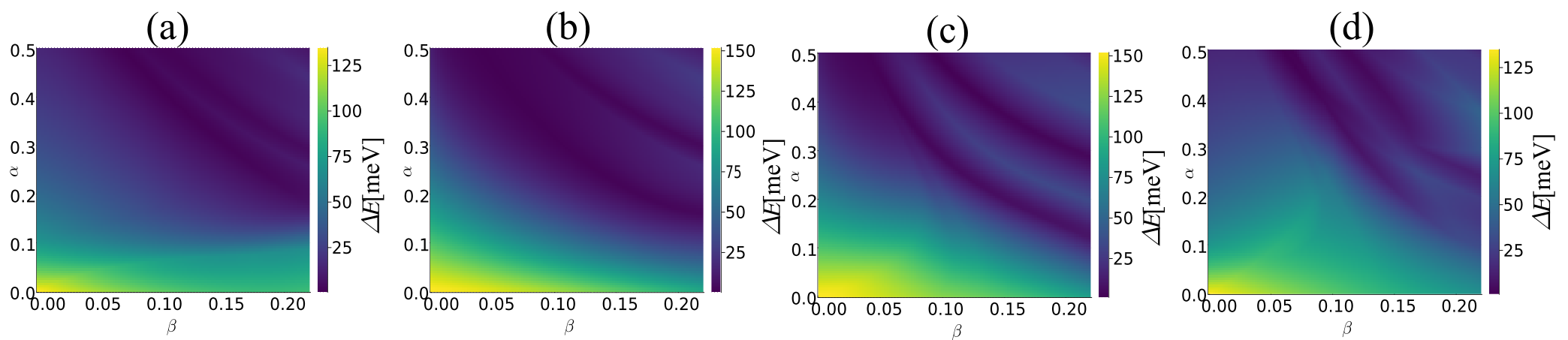}
    \caption{The band width of the each band of PSMG as a function of dimensionless parameters $\alpha$ and $\beta$.(a-d) for band index $-1,0,1,2$ respectively.}
    \label{fig:PSMG_band_width}
\end{figure}

Considering the significance of band geometry in the formation of FCIs, we also calculate the Berry curvature $B\left(\vb*{k}\right)$ and Fubini-Study (F-S) metric $g^{ab}\left(\vb*{k}\right)$ of PSMG. We show the fluctuation of Berry curvature and F-S metric of each band of PSMG
as a function of dimensionless parameters $\alpha$ and $\beta$ in [Fig. \ref{fig:PSMG_fluctuation}], where the fluctuations of Berry curvature and F-S metric are quantified by
\begin{equation}
    \sigma \left( B \right)=\sqrt{\left\langle {{B}^{2}}\left( {\vb*{k}} \right) \right\rangle -{{\left\langle B\left( {\vb*{k}} \right) \right\rangle }^{2}}},
\end{equation}
\begin{equation}
    \sigma \left( g \right)=\sqrt{\sum\limits_{a,b}{\left( \left\langle {{g}^{ab}}{{g}^{ba}} \right\rangle -\left\langle {{g}^{ab}} \right\rangle \left\langle {{g}^{ba}} \right\rangle  \right)}},
\end{equation}
where $\left\langle O\left( \mathbf{k} \right) \right\rangle =\frac{1}{{{A}_{BZ}}}\int\limits_{BZ}{\dd[2]{k}O\left( \mathbf{k} \right)}$. Regions with small fluctuations of Berry curvature and F-S metric are candidates for realizing FCIs.
\begin{figure}[htbp]
    \centering
    \includegraphics[width=1.0\textwidth]{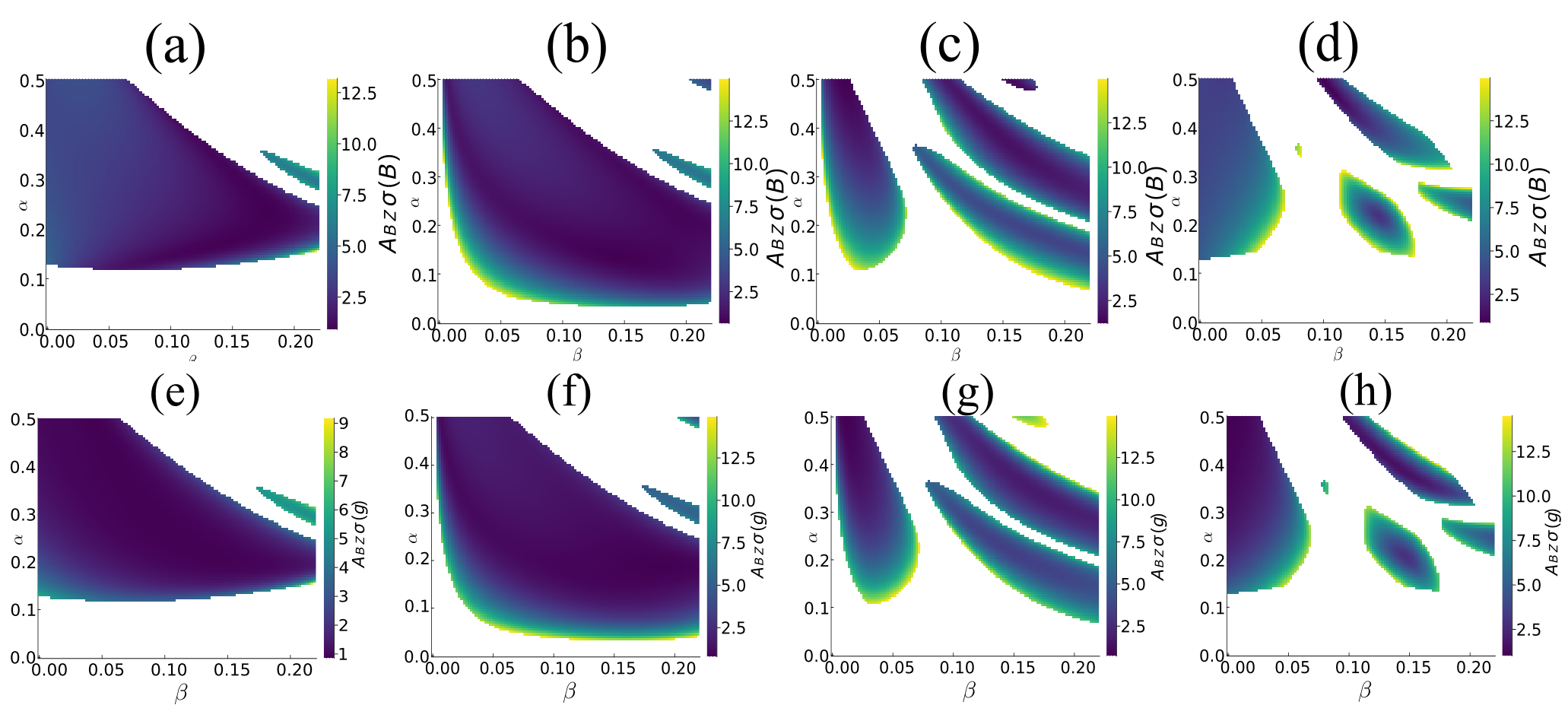}
    \caption{The fluctuation of Berry curvature and F-S metric of each band of PSMG as a function of dimensionless parameters $\alpha$ and $\beta$. Only regions with band gap larger than $6\text{meV}$ and Berry curvature fluctuation smaller than $15/A_{BZ}$ and F-S metric fluctuation smaller than $15/A_{BZ}$ are shown, where $A_{BZ}$ is the area of mini-Brillouin zone. (a-d) for Berry curvature fluctuation of band index $-1,0,1,2$ respectively. (e-h) for F-S metric fluctuation of band index $-1,0,1,2$ respectively.}
    \label{fig:PSMG_fluctuation}
\end{figure}

\section{ADDITIONAL EVIDENCE OF Laughlin FCIs IN PSMG $\left| \mathcal{C} \right|=1$ BANDS}
\label{sec:additional_spectra}

\begin{figure}[htbp]
    \centering
    \includegraphics[width=1.0\textwidth]{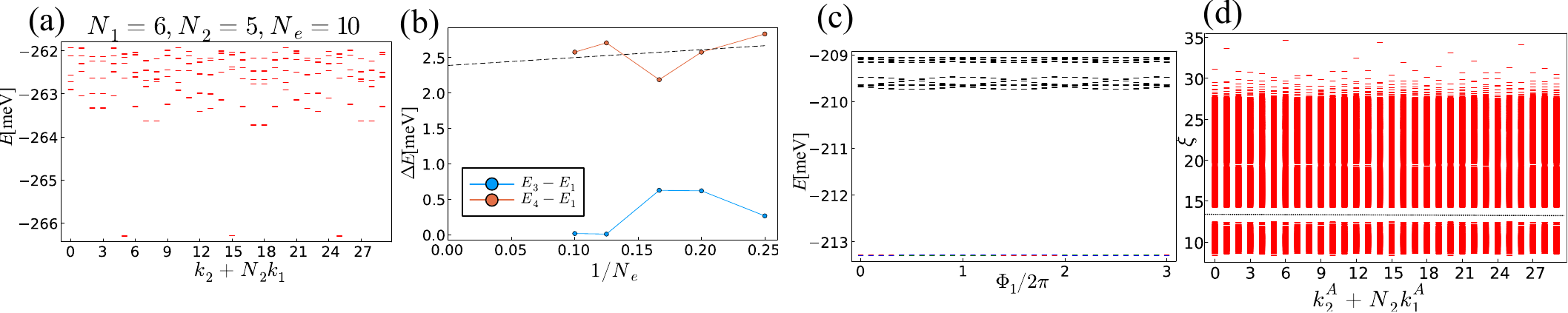}
    \caption{Evidence of $\nu=1/3$ Laughlin FCIs in PSMG. (a) Low-lying energy spectra for a $6\times5$ system.
    (b) Finite-size scaling of the energy gap $E_4-E_1$ (orange) and the ground-state splitting $E_3-E_1$ (blue) for systems with $N_e=4\sim10$ electrons.
    (c) Energy spectral flow with magnetic flux $\Phi_1$ insertion in the $\vb*{a}_1$ direction for a $6\times4$ system.
    (d) Particle entanglement spectrum for $N_A=5$ in a $6\times5$ system, showing 23256 levels below the entanglement gap (dashed line). Here, we consider representative parameter $\alpha=0.4,\beta=0.068$.}
    \label{fig:PSMG_alpha_beta}
\end{figure}

In this part, we provide additional evidence for the realization of Laughlin FCIs in PSMG $\left| \mathcal{C} \right|=1$ bands.
For parameter point $\alpha=0.4,\beta=0.068$ in region $A_{\left(0\right)}$, we perform larger system size calculations assuming spin polarization (see Fig. \ref{fig:PSMG_alpha_beta}(a)), there is a three-fold (quasi-)degeneracy in the low-lying energy spectra, and their corresponding momenta align with the predictions of Haldane statistics for $\nu=1/3$ Laughlin FCI.
We further perform finite-size scaling of the many-body gap $E_4-E_1$ and the ground-state splitting $E_3-E_1$, here all energy eigenvalues ${E_n}$ are sorted in ascending order. As illustrated in Fig.~\ref{fig:PSMG_alpha_beta}(b), $E_4-E_1$ approximately saturates to a finite value ($\sim$2.4meV) while the ground-state splitting vanishes, demonstrating 
robust ground-state degeneracy~\cite{laughlin_anomalous_1983} towards the thermodynamic limit.
To further confirm the ground-state topology, we examine both the energy spectra flow by inserting magnetic flux and the particle-cut entanglement spectrum (PES). Figure \ref{fig:PSMG_alpha_beta}(c) exhibits the spectral flow with magnetic flux inserting along $\vb*{a}_1$ direction, the persistence of the three-fold degeneracy corroborates the topological nature of ground states. The PES of the ground states is computed by partitioning the system into $N_A$ and $N_e-N_A$ electrons. The PES exhibits an entanglement gap separating the low-lying levels from higher ones [see Fig.~\ref{fig:PSMG_alpha_beta}(d)], and the level counting below this gap matches the quasihole excitations in the $\nu=1/3$ Abelian FCIs~\cite{sterdyniak_extracting_2011,li_entanglement_2008}. This evidence rules out competing states such as charge density waves (CDWs) and further verifies the ground-state topology.
These findings strongly support the existence of $\nu=1/3$ Laughlin FCIs in PSMG. We remark that our results of energy spectra at $\nu_T=-4+1/3$ ($\nu_T=-4+\nu$) are consistent with Ref. \cite{gao_untwisting_2023} for $\nu_T=-2/3$. Moreover, we identify the ground-state topology by PES [see Figs.~\ref{fig:PSMG_alpha_beta}(d)].

Assuming spin-polarization, we also provide many-body spectra of various strain parameters supporting the realization of FCIs in PSMG that are not included in the main text. 
We show the low-lying energy spectrum of some representative parameter points for regions $B_{\left(0\right)}$, $C_{\left(1\right)}$, $D_{\left(1\right)}$, and $E_{\left(2\right)}$ in Fig. \ref{fig:PSMG_additional_spectra}. When these bands are $1/3$ filled, the three-fold ground-state degeneracy and the robust many-body energy gap are observed in all these regions.
These regions correspond to band Chern number $\left|{\mathcal{C}}\right|=1$, so we infer that the $\nu=1/3$ Laughlin state is likely to form in these regions.

\begin{figure}[htbp]
    \centering
    \includegraphics[width=1.0\textwidth]{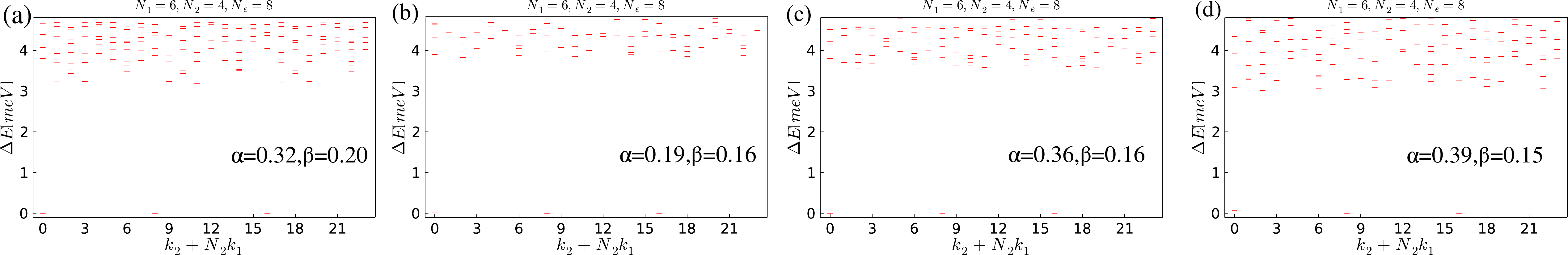}
    \caption{The low-lying energy spectrum for some representative parameter points for regions $B_{\left(0\right)}$, $C_{\left(1\right)}$, $D_{\left(1\right)}$, and $E_{\left(2\right)}$ with band Chern number $\left|\mathcal{C}\right|=1$ in PSMG.}
    \label{fig:PSMG_additional_spectra}
\end{figure}
\section{Nonrelativistic Electrons in a Periodic Magnetic Field}

It has been established that a nonrelativistic two-dimensional electron gas in a periodic magnetic field, governed by the Hamiltonian
\begin{equation}
    \mathcal{H}=\frac{{{\hbar }^{2}}}{2m}{{\left( -i\nabla -\frac{e}{\hbar} \vb*{A} \right)}^{2}},
\end{equation}
can give rise to a series of topologically nontrivial minibands~\cite{taillefumier_chiral_2008}. Using the same magnetic field and effective mass parameters as in the main text for PSTMDs, we explore the band geometry of these minibands. Our analysis reveals that the deviation from the trace condition,
\begin{equation}
    \delta_{\operatorname{tr}}=\left\langle \operatorname{tr}g\left( {\vb*{k}} \right)-\left| B\left( {\vb*{k}} \right) \right| \right\rangle,
\end{equation}
for the lowest band which corresponds to Chern number of 1, can be remarkably small when $\gamma \equiv -\frac{e{{A}_{0}}}{\hbar \left| {{G}_{0}} \right|} \approx 0.22$ (see Fig. \ref{fig:PSTMD_nonrelativistic_band}), indicating the potential realization of Fractional Chern Insulators (FCIs) in this system. To confirm the presence of FCIs, we perform exact diagonalization (ED) on the projected Hamiltonian of the lowest band at $\nu=1/3$ filling, using the same screened Coulomb potential as described in the main text (see Fig. \ref{fig:PSTMD_nonrelativistic_band}).

\begin{figure}
    \centering
    \includegraphics[width=1.0\textwidth]{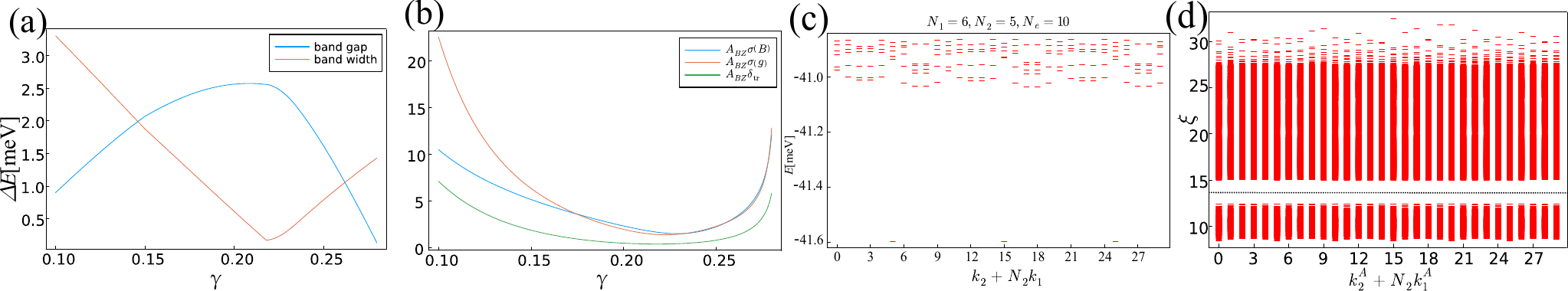}
    \caption{(a) Bandwidth and indirect band gap of the lowest band as a function of $\gamma$. 
    (b) Fluctuation of Berry curvature, Fubini-Study metric, and deviation from the trace condition of the lowest band as a function of $\gamma$.
At $\gamma = 0.22$ and $\nu = 1/3$ in a $6 \times 5$ system, panels (c-d) show the low-lying energy spectra (c) and particle entanglement spectrum (d) for $N_A = 5$ with 23,256 levels below the entanglement gap. }
    \label{fig:PSTMD_nonrelativistic_band}
\end{figure}

\section{spin polarization evolution with filling factor}

In the main text, we have shown spin polarization evolution with filling factor in nearly ideal $\mathcal{C}=1$ band. While a rigorous theory of spin polarization in FCIs as a function of filling factor is still under development, we present a qualitative argument based on the energy minimization principle.

At lower filling factors, such as $\nu=1/3$, the system tends to form a fully spin-polarized state. This tendency can be understood in terms of the interplay between electron-electron repulsive (Coulomb) interactions and wavefunction symmetry. At low filling factors, electrons occupy states sparsely within the Landau-like bands, leading to relatively large average distances between electrons. The Coulomb repulsion thus plays a dominant role in determining the system's ground state. Since electrons are fermions, the total wavefunction must be antisymmetric upon exchange. For a symmetric spin state (i.e., fully spin-polarized), the spatial part of the wavefunction must be antisymmetric, maximizing inter-electron distances and reducing Coulomb energy. Thus, spin polarization is favorable at low filling factors as it minimizes the total energy.

At higher filling factors, such as $\nu=2/3$, electron density is higher, reducing the average inter-electron distance and thus intensifying Coulomb interactions. While spin polarization reduces Coulomb energy to some extent, when electrons are closely spaced, the repulsive interaction can outweigh the energy gains from spin polarization. In this regime, the system may further reduce its energy by transitioning to an unpolarized or partially polarized state. Such states allow electrons to occupy both spin subspaces, increasing spatial freedom within the Landau-like levels and thus decreasing overall repulsion.
 
\section{Evidence for Spin Polarization and Spin Unpolarization in PSMG}

Due to limited computational resources, we can only compute the energy spectra of the spinful Hamiltonian on relatively small system sizes. This limitation leads to a relatively large energy splitting among the quasi-degenerate ground states at $\nu=2/3$ and $\nu=2/5$ presented in the main text. To confirm the robustness of these states, we provide further evidence here.

For $\nu=2/3$, we present the energy spectra for different system sizes [see Fig.~\ref{112_Halperin_scaling}(a-c)], which show qualitative agreement. Additionally, we have computed the spectral flow under the insertion of magnetic flux through the handle of the torus [see Fig.~\ref{112_Halperin_scaling}(d)]. We observe that the FCI ground states evolve into each other during the spectral flow with appropriate flux insertion, without mixing with higher excited levels. This indicates that the energy spectrum remains qualitatively consistent under different boundary conditions, reducing the likelihood that finite-size effects would significantly alter the nature of the ground state.

\begin{figure}[htbp]
    \begin{center}
    \includegraphics[width=1.0\textwidth]{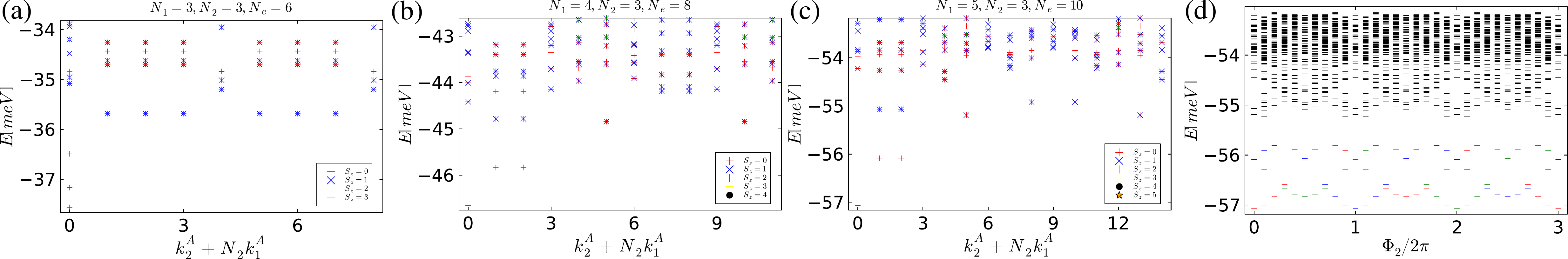}
    \end{center}
    \caption{Evidence for the Halperin (112) state at $\nu=2/3$ in PSMG. Energy spectra for different $S_z$ sectors and system sizes (from $N_e=6$ to $N_e=10$) are shown in (a-c). The spectral flow under magnetic flux insertion through the handle of the torus is shown in (d).}
    \label{112_Halperin_scaling}
\end{figure}

For $\nu=2/5$, we present the energy spectra for two different system sizes [see Fig.~\ref{332_Halperin_scaling}(a-b)], which also exhibit qualitative agreement. For both the $N_e=6$ and $N_e=8$ systems, our calculations show that the ground-state energy splitting is significantly smaller than the energy gap (denoted as $\Delta$) between the quasi-degenerate ground states and the excited states. Notably, the value of $\Delta$ increases slightly when the system size is increased from $N_e=6$ to $N_e=8$.

\begin{figure}[htbp]
    \begin{center}
    \includegraphics[width=1.0\textwidth]{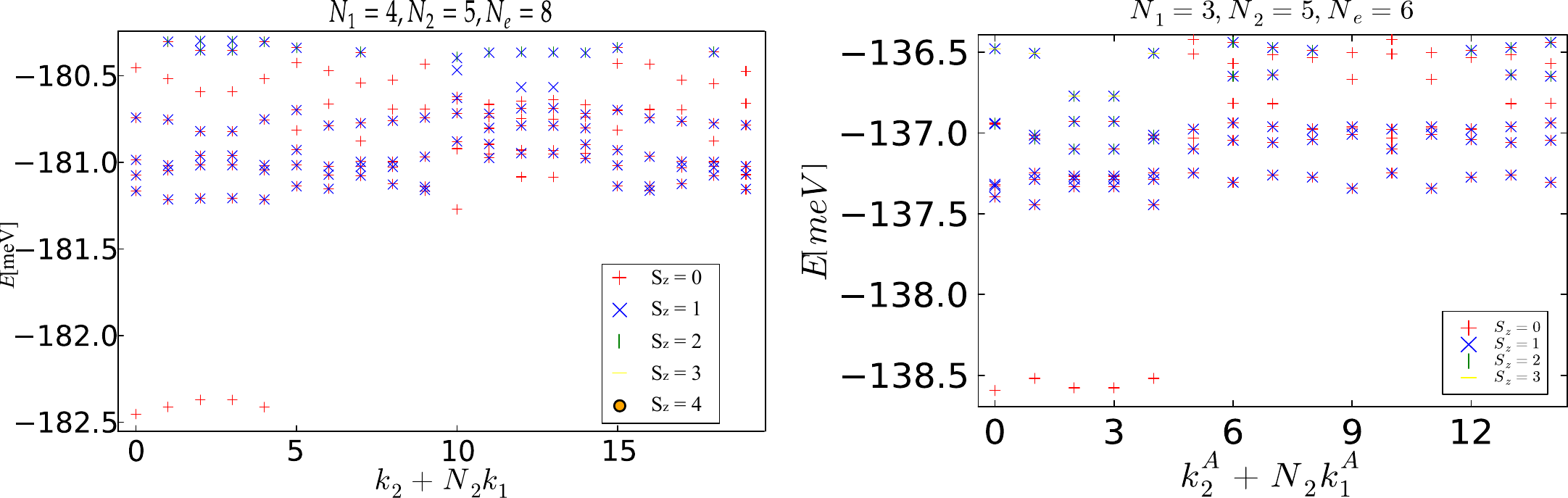}
    \end{center}
    \caption{Evidence for the Halperin (332) state at $\nu=2/5$ in PSMG. Energy spectra for different $S_z$ sectors and system sizes (from $N_e=6$ to $N_e=8$) are shown.}
    \label{332_Halperin_scaling}
\end{figure}

This part we also offer ED results supporting the realization of spin-polarized ground states in PSMG that are not included in the main text.

At $\nu=2/3$ filling of the region $B_{\left(-1\right)}$,  $\nu=1/3$ filling of the region $D_{\left(2\right)}$ and $\nu=1$ filling of the region $A_{\left(0\right)}$, we find ground state energies with different $S_z$ are identical [Fig.\ref{fig:polarized_states}, Fig. \ref{fig:exciton_superfluid}(a)], indicating a ferromagnetic ground state with total spin $S = N_e/2$. 
\begin{figure}[htbp]
    \centering
    \includegraphics[width=1.0\textwidth]{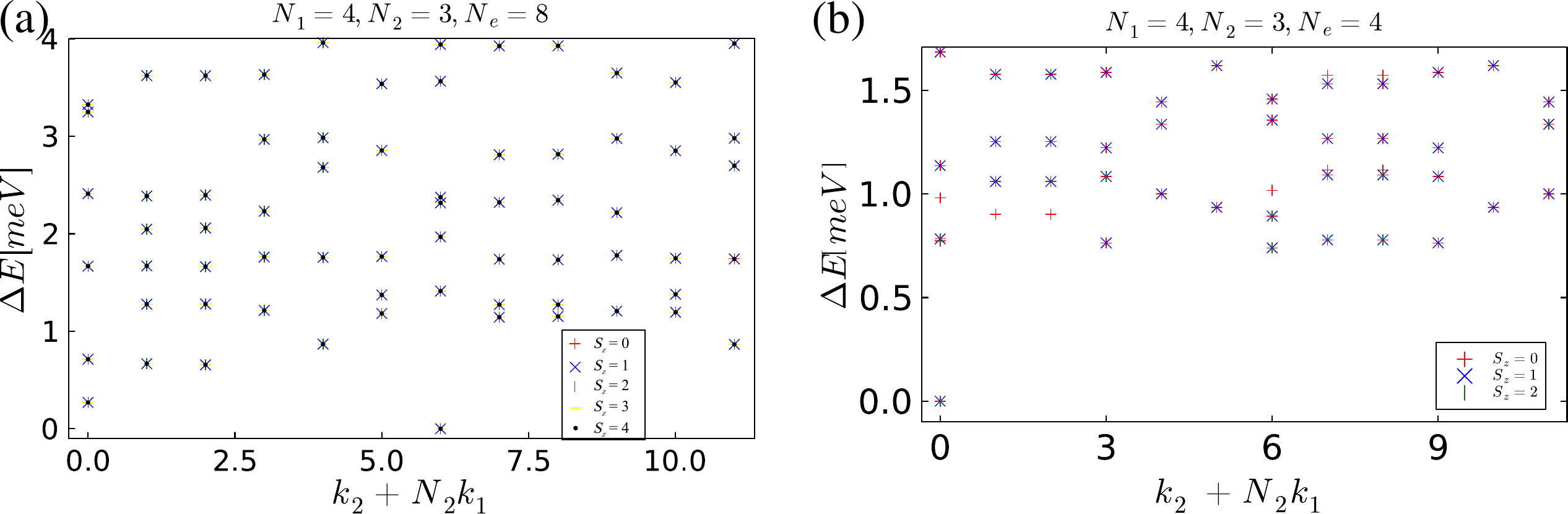}
    \caption{Evidence of spin-polarized states in $\left|\mathcal{C}\right|=2$ bands of PSMG. (a) The low-lying energy spectrum for different $S_z$ at $\nu=2/3$ filling of the region $B_{\left(-1\right)}$, with $\alpha = 0.29,\beta=0.22$. (b) The low-lying energy spectrum for different $S_z$ at $\nu=1/3$ filling of the region $D_{\left(2\right)}$, with $\alpha = 0.29,\beta=0.125$.}
    \label{fig:polarized_states}
\end{figure}
For [Fig.\ref{fig:polarized_states}], though quasi-degenerate ground states are unobservable due to finite-size effects, these results show that the spin-polarized state is likely to be the ground state or at least energetically competitive with other states. Even if the ground state is not fully spin-polarized, with weak magnetic fields, the system is likely to be spin-polarized in the thermodynamic limit.

Moreover, for $\nu=1$ filling of the region $A_{\left(0\right)}$, we examine the in-plane spin correlations and the energy flow with twisted boundary conditions, both indicate that such a ferromagnetic state is more consistent with an $XY$-type easy-plane ferromagnet rather than an Ising ferromagnet. 
We compute the spin correlation function $\left\langle S_{-\vb*{q}}^{y}S_{{\vb*{q}}}^{y} \right\rangle $ for the ground state with $S_z=0$ and find a clear peak at $\vb*{q}=\vb*{0}$ [Fig. \ref{fig:exciton_superfluid}(b)], suggesting the ferromagnetism in the $XY$ easy-plane. 
With twisted (opposite) boundary conditions, the ground state energy  varies as:
\begin{equation}
    E\left( {{\theta }_{t}} \right)/A=E\left( {{\theta }_{t}}=0 \right)/A+\frac{1}{2}{{\rho }_{s}}\theta _{t}^{2}+\mathcal{O}\left( \theta _{t}^{4} \right),
\end{equation}
where $\theta_t$ is the twist angle, $A$ is the area of the system, and $\rho_s$ is the stiffness. We observe this parabolic behavior in the ground state energy [Fig. \ref{fig:exciton_superfluid}(c)] with finite spin stiffness, which suggests the spin channel condenses and becomes a gapless spin superfluid. These observations are consistent with the properties of the two-component Halperin (111) state in the isotropic limit (i.e., the same inter- and intra-component interactions). Here, ``two-component" denotes spin degrees of freedom, which also could represent general labels including pseudo-spin, layer, valley, sub-band, etc., in the context of bilayer quantum Hall issue.

\begin{figure}[htbp]
    \centering
    \includegraphics[width=1.0\textwidth]{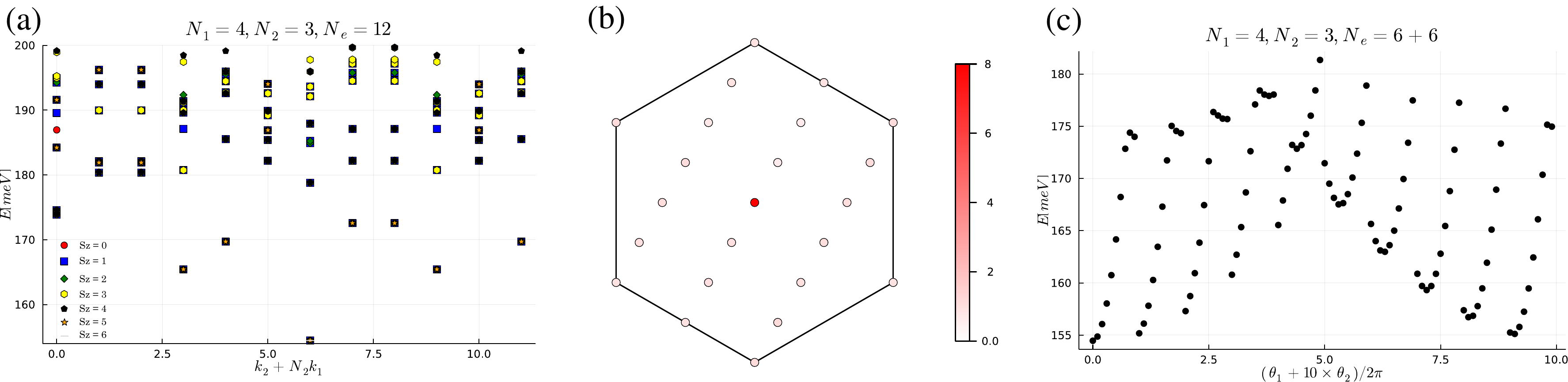}
    \caption{Evidence of Halperin (111) state in PSMG with $\alpha=0.4,\beta=0.068$. (a) The ground state energy for different $S_z$ at $N_1=4,N_2=3,N_e=12$. (b) The spin correlation function $\left\langle S_{-\vb*{q}}^{y}S_{{\vb*{q}}}^{y} \right\rangle $ for the ground state with $S_z=0$. (c) The ground state energy as a function of twist angle $\theta_t$ for $S_z=0$.}
    \label{fig:exciton_superfluid}
\end{figure}

\section{Density correlation and occupation number of the Laughlin-like state in PSMG $\left| \mathcal{C} \right|=2$ band}
To further understand the nature of the the Laughlin-like state in PSMG $\left| \mathcal{C} \right|=2$ band, we study its
projected density structure factor and occupation number of the ground state in Fig. \ref{fig:PSMG_high_C}(b) of the main text. The projected density structure factor is defined as
\begin{equation}
    S\left( \mathbf{q} \right)=\frac{1}{{{N}_{1}}{{N}_{2}}}\left( \left\langle {{{\bar{\rho }}}_{\mathbf{q}}}{{{\bar{\rho }}}_{-\mathbf{q}}} \right\rangle -{{N}_{e}}^{2}{{\delta }_{\mathbf{q},0}} \right),
    \label{eq:projected_density_structure_factor}
\end{equation}
where $\bar{\rho}_{\mathbf{q}}$ is the projected density operator in the momentum space. As shown in Fig. \ref{fig:PSMG_Laughlin_corrlation}(b), the projected density structure factor exhibits no peak at the whole Brillouin zone, indicating that the ground state does not break the translational symmetry. 

The occupation number of the ground state is shown in Fig. \ref{fig:PSMG_Laughlin_corrlation}(a). We find the occupation number in $\Gamma$ point almost vanishes. The scattering processes associated with the $\mathbf{k}=0$ single particle state are invisible by the ground state, which supports our argument for the stabilization of the Laughlin-like state in the $\left|\mathcal{C}\right|=2$ band in the main text.

\begin{figure}
    \centering
    \includegraphics[width=1.0\textwidth]{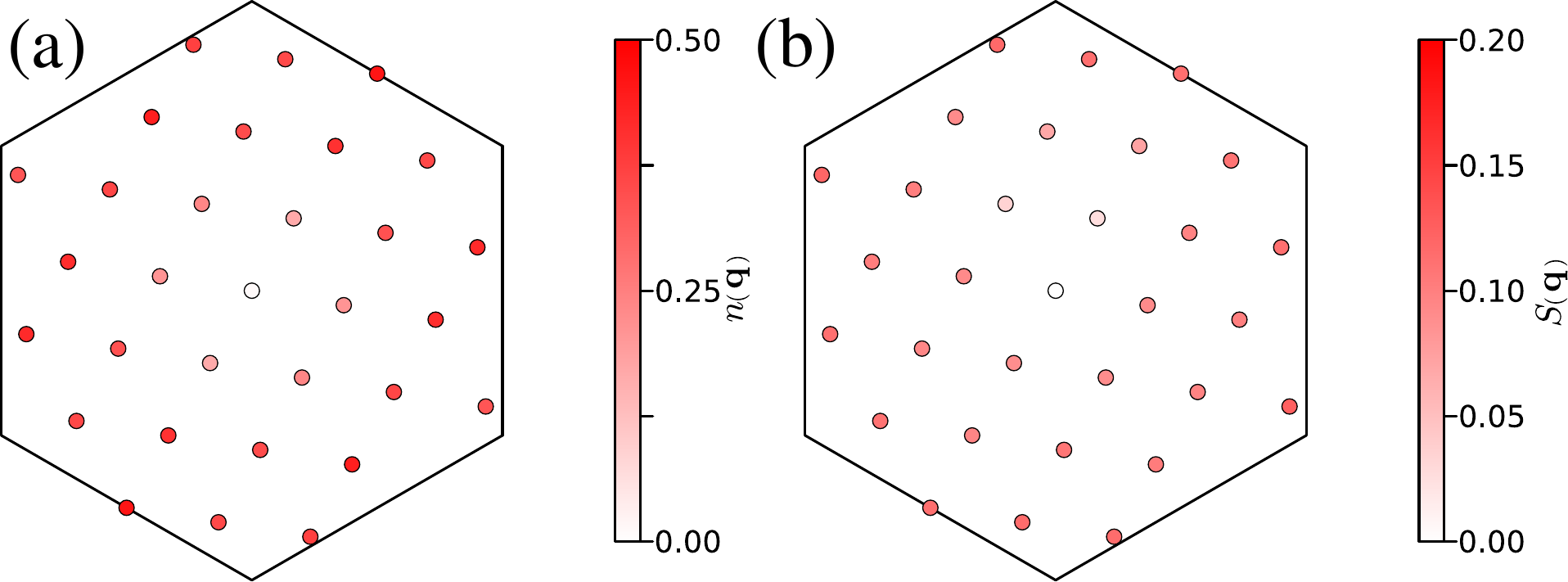}
    \caption{Projected density structure factor and occupation number of the Laughlin-like state in PSMG $\left| \mathcal{C} \right|=2$ band. (a) The occupation number of the ground state at $\nu=1/3$ filling of the region $D_{\left(2\right)}$.(b) The projected density structure factor $S\left( \mathbf{q} \right)$ for the ground state at $\nu=1/3$ filling of the region $D_{\left(2\right)}$.}
    \label{fig:PSMG_Laughlin_corrlation}

\end{figure}

\end{widetext}

\end{document}